\documentclass[twocolumn,showpacs]{revtex4}
\usepackage{mathrsfs}
\usepackage{multirow}
\usepackage{longtable,lscape}
\usepackage{subfigure}
\usepackage{float}
\usepackage{amssymb}
\usepackage{indentfirst}
\usepackage{graphicx,booktabs}
\usepackage{color}
\usepackage{epstopdf}
\usepackage{txfonts}

\def\beq{\begin{equation}}
\def\enq{\end{equation}}
\def\beqa{\begin{eqnarray}}
\def\enqa{\end{eqnarray}}
\def\nn{\nonumber}
\def\lb{\label}
\def\MeV{\nobreak\,\mbox{MeV}}
\def\GeV{\nobreak\,\mbox{GeV}}
\def\al{\alpha}
\def\be{\beta}
\def\almax{\alpha_{max}}
\def\almin{\alpha_{min}}
\def\bemax{1-\al}
\def\bemin{\beta_{min}}
\newcommand{\rag}{\rangle}
\newcommand{\lag}{\langle}
\def\qqqq{\lag\bar{q}q\bar{q}q\rag}
\def\qq{\langle \bar{q} q \rangle}
\def\qGq{\lag\bar{q}Gq\rag}
\def\GG{\lag G^2 \rag}
\def\gG{\lag g_s^2 G^2 \rag}
\def\GGG{\lag G^3 \rag}
\def\gGGG{\lag g_s^3 G^3 \rag}
\def\8{\lag 8 \rag}

\begin{document}

\title{Exotic $B_c$-like molecules in QCD sum rules}

\author{Raphael M. Albuquerque}
\email{rma@if.usp.br}
\affiliation{Instituto de F\'{\i}sica, Universidade de S\~{a}o Paulo,
C.P. 66318, 05315-970 S\~{a}o Paulo, SP, Brazil}

\author{Xiang Liu$^{1,2}$}
\email{xiangliu@lzu.edu.cn}
\affiliation{ $^1$School of Physical Science and Technology, Lanzhou University,
Lanzhou 730000, China\\
$^2$Research Center for Hadron and CSR Physics,
Lanzhou University and Institute of Modern Physics of CAS, Lanzhou 730000, China}

\author{Marina Nielsen}\email{mnielsen@if.usp.br}
\affiliation{Instituto de F\'{i}sica, Universidade de S\~{a}o
Paulo, C.P. 66318, 05315-970 S\~{a}o Paulo, SP, Brazil}


\begin{abstract}
We use the QCD  sum rules to study possible $B_c$-like molecular states.
We consider isoscalar $J^P=0^+$ and $J^P=1^+$ $D^{(*)}B^{(*)}$ molecular currents.
We consider the contributions of condensates up to dimension eight and we work at
leading order in $\alpha_s$. We obtain for these states masses  around 7 GeV.
\end{abstract}

\pacs{14.40.Rt, 12.39.Pn, 13.75.Lb} \maketitle


The study of states with configuration more complex than the
conventional $q\bar{q}$ meson and $qqq$ baryon is quite old and, despite decades
of progress, no exotic hadron has been conclusively identified.
Famous examples of possible nonconventional meson states are the light scalars
 and the $X(3872)$ \cite{Choi:2003ue}. While, from a theoretical point of
view the most acceptable structure for the light scalars is a tetraquark
(diquark-antidiquark) configuration \cite{jaffe}, in the case of the $X(3872)$
there is an agreement in the community that it might be a $D\bar{D}^*$ molecular state.
Establishing the structure of these states and identifying other
possible exotic states represents a remarkable progress in hadron physics.

Besides the $X(3872)$, in the past decade, more and more
charmonium-like or bottomonium-like states were observed in the
$e^+e^-$ collision \cite{Aubert:2005rm,Aubert:2006ge,:2007sj}, $B$
meson decays \cite{Choi:2003ue,Abe:2004zs,:2007wga,Aaltonen:2009tz} and
even $\gamma\gamma$ fusion processes \cite{Uehara:2005qd,Uehara:2009tx,Shen:2009vs},
which have stimulated the extensive discussion of exotic hadron configurations
(for a review see Refs.
\cite{Swanson:2006st,Zhu:2007wz,Nielsen:2009uh,Brambilla:2010cs}).
An important question that arrises is that if some of these observed
states are molecular states, then many others should also exist. In a
very recent publication \cite{Sun:2012sy}, a one boson exchange (OBE) model
was used to investigate hadronic molecules with both open
charm and open bottom. These new structures were
labelled as  $B_c$-like molecules, and were categorized into four
groups: $\mathcal{DB}$, $\mathcal{D}^*\mathcal{B}^*$,
$\mathcal{D}^*\mathcal{B}$ and $\mathcal{D}\mathcal{B}^*$, where these symbols
represent the group of states: $\mathcal{D}^{(*)}=[D^{(*)0},D^{(*)+},D^{(*)+}]$
for charmed mesons and $\mathcal{B}^{(*)}=[B^{(*)+},B^{(*)0},B^{(*)0}]$ for
bottom mesons. A complete analysis, based on the approach developed in Refs.
\cite{Tornqvist:1993vu,Tornqvist:1993ng,Swanson:2003tb,Liu:2008fh,Liu:2008tn,Thomas:2008ja,Lee:2009hy,Sun:2011uh},
was done in Ref.~\cite{Sun:2012sy} to study the interaction of these $B_c$-like
molecules. These states were categorized using a hand-waving notation, with
five-stars, four-stars,
etc. A five-star state implies that a loosely molecular state probably exists. They
find five five-star states, all of them isosinglets in the light sector, with no
strange quarks.

Here we use the QCD sum rules (QCDSR) \cite{Nielsen:2009uh,svz,rry,SNB}, to
check if some of the five-star states found in Ref.~\cite{Sun:2012sy} are
supported by a QCDSR calculation. The states we will consider are the isosinglets
$J^P=0^+$ $DB=(D^0B^++D^+B^0)$, $J^P=1^+$ $D^*B=(D^{*0}B^++D^{*+}B^0)$,
$J^P=1^+$ $DB^*=(D^{0}B^{*+}+D^{+}B^{*0})$ and the $J^P=0^+$ $D^*B^*=(D^{*0}
B^{*+}+D^{*+}B^0)$.
The QCDSR approach is based on the  two-point correlation function
\beq
  \Pi(q)=i\int d^4x ~e^{iq.x}\lag 0| T [j(x)j^\dagger(0)] |0\rag,
  \lb{2po}
\enq
where the current $j(x)$ contains all the information about the hadron of
interest, like quantum numbers, quarks contents and so on. Possible currents
for the states described above are given in Table \ref{mol}, where
we have used a short notation for the isoscalars since we
are considering the light quarks, $q=u,~d$, degenerate. We use the same
techniques developed in  Refs.~
\cite{Albuquerque:2009ak,x3872,molecule,lee,bracco,rapha,z12,zwid,mix,x4350,Finazzo:2011he,Albuquerque:2011ix}.

{\small
\begin{table}[h]
\setlength{\tabcolsep}{1.25pc}
\caption{Currents describing possible $B_c$-like molecules.}
\begin{tabular}{ccl}
&\\
\hline
State &$I(J^{P})$ & Current\\
\hline
$D\:B$ & $0(0^{+})$ & $ j = (\bar{q} \gamma_5 c)(\bar{b} \gamma_5 q)$ \\
$D^\ast B^\ast$ & $0(0^{+})$ & $j=(\bar{q} \gamma_\mu c)(\bar{b}\gamma^\mu q)$ \\
$D^\ast B$ & $0(1^{+})$ & $ j_\mu =i(\bar{q} \gamma_\mu c)(\bar{b}\gamma_5 q)$ \\
$D B^*$ & $0(1^{+})$ & $ j_\mu = i(\bar{q} \gamma_5 c)(\bar{b}\gamma_\mu q)$ \\
\hline
\end{tabular}
\label{mol}
\end{table}}

The QCD sum rule is obtained by evaluating the correlation function in
Eq.~(\ref{2po}) in two ways: in the OPE side, we calculate the correlation
function at the quark level in terms of quark and gluon fields.  We work at
leading order in $\alpha_s$ in the operators, we consider the contributions
from condensates up to dimension eight. In the phenomenological side,
the correlation function is calculated by inserting intermediate states
for the hadronic state, $H$, and parameterizing the coupling of these states to the
current $j_\mu(x)$, in terms of a generic coupling parameter $\lambda$, so that:
\beq
  \lag 0 | j| H \rag =  \lambda ,
  \lb{o+}
\enq
for the scalar states and
\beq
  \lag 0 | j_\mu| H \rag =  \lambda \:\varepsilon_\mu,
  \lb{1+}
\enq
for the axial currents, where $\varepsilon_\mu$ is the polarization vector.
In the case of the axial current, we can write the correlation function
in Eq.~(\ref{2po}) in terms of two independent Lorentz structures:
\beq
\Pi_{\mu\nu}(q)=-\Pi(q^2)(g_{\mu\nu}-{q_\mu q_\nu
\over q^2})+\Pi^\prime(q^2){q_\mu q_\nu\over q^2}.
\lb{lorentz}
\enq
The two invariant functions, $\Pi$ and $\Pi^\prime$, appearing in
Eq.~(\ref{lorentz}), have respectively the quantum numbers of the spin 1
and 0 mesons. Therefore, we choose to work with the Lorentz structure
$g_{\mu\nu}$, since it projects out the $1^{+}$ state.

The phenomenological side of Eq.~(\ref{2po}), in the $g_{\mu\nu}$ structure
in the case of the axial currents, can be written as
\beq
\Pi^{phen}(q^2)={\lambda^2\over M_{_H}^2-q^2} + \int\limits_{0}^\infty ds\,
{\rho^{cont}(s)\over s-q^2},
\lb{phe}
\enq
where $M_{_H}$ is the hadron mass and the second term  in the RHS
of Eq.~(\ref{phe}) denotes the contribution
of the continuum of the states with the same quantum numbers as the current.
In general, in the QCDSR method it is
assumed that the continuum contribution to the spectral density,
$\rho^{cont}(s)$ in Eq.~(\ref{phe}), vanishes below a certain continuum
threshold $s_0$. Above this threshold, it is given by
the result obtained in the OPE side. Therefore, one uses the ansatz \cite{io1}
\beq
  \rho^{cont}(s)=\rho^{OPE}(s)\Theta(s-s_0)\;.
\enq

The correlation function in the OPE side can be written as a
dispersion relation:
\beq
  \Pi^{OPE}(q^2)=\int_{(m_c+m_b)^2}^\infty ds {\rho^{OPE}(s)\over s-q^2}\;,
  \lb{ope}
\enq
where $\rho^{OPE}(s)$ is given by the imaginary part of the
correlation function: $\pi \rho^{OPE}(s)=\mbox{Im}[\Pi^{OPE}(s)]$.

After transferring the continuum contribution to the OPE side, and
performing a Borel transform, the sum rule can be written as
\beq
  \lambda^2e^{-M^2_{_H} \:\tau}=\int\limits_{(m_c+m_b)^2}^{s_0}ds~
  e^{-s \:\tau}~\rho^{OPE}(s)\;,
  \label{sr1}
\enq
where we have introduced the Borel parameter $\tau=1/M^2$, with $M$ being the
Borel mass. To extract $M_{_H}$ we take the derivative of Eq.~(\ref{sr1}) 
with respect to Borel parameter $\tau$ and divide the result by Eq.~(\ref{sr1}), 
so that:
\beq
  M^2_{_H} = \frac{\int\limits_{4m_Q^2}^{s_0}ds~ s \:e^{-s \:\tau}~\rho^{ope}(s)}
  {\int\limits_{4m_Q^2}^{s_0}ds~ e^{-s \:\tau}~\rho^{ope}(s)}\;.
  \label{mass}
\enq

The expressions for $\rho^{ope}(s)$ for the currents in Table \ref{mol},
using factorization hypothesis, up to dimension-eight condensates, are given  in
appendix \ref{App1}.

To extract reliable results from the sum rule, it is necessary to establish the
Borel window. A valid sum rule exists when one can find a Borel window where
 there are a OPE  convergence, a $\tau$-stability and the dominance
of the pole contribution. The maximum value of $\tau$ parameter is determined by
imposing that the contribution of the higher dimension condensate is smaller
than 15\% of the total contribution. The minimum value of $\tau$ is determined
by imposing that the pole contribution is equal to the continuum
contribution. To guarantee a reliable result extracted from sum rules it is
important that there is a $\tau$ stability inside the Borel window.

The continuum threshold is a physical parameter that should be determined from
the spectrum of the mesons. Using a harmonic-oscillator potential model, it was 
shown in Ref.~\cite{Lucha:2007pz} that a constant continuum threshold is a 
very poor approximation. The actual accuracy of the parameters extracted from 
the sum rules improves considerably when using a Borel dependent continuum 
threshold. It also allows to estimate realistic systematic errors 
\cite{Lucha:2007pz}. However, to be able to fix the form of the Borel dependent 
continuum threshold (and the values of the parameters in the function) one needs 
to use the experimental value of the mass of the particle \cite{Lucha:2011zp}. 
Since in our study we do not know the experimental value of the masses of the 
states, it is not possible to fix the Borel dependent continuum threshold.
For this reason, although aware of the limitations of the values we are going to
extract from the sum rule, to have a first estimate for the values of the 
masses of the states, we are going to use a constant continuum threshold.
In many cases, a good approximation for the value of the continuum threshold
is the value of the mass of the first excited
state squared. In some known cases, like the $\rho$ and $J/\psi$,
the first excited state has a mass approximately $0.5~\GeV$ above the
ground state mass.  Since here we do not know the spectrum for the 
hadrons studied, we will 
fix the continuum threshold range starting with the smaller value which 
provides a valid Borel window. The optimal choice for
$s_0$ will be taken when there is a$\tau-$stability inside the Borel window.

For a consistent comparison with the results obtained for the other molecular
states using the QCDSR approach, we have considered here the same values
used for the quark masses and condensates as in
Refs.~\cite{x3872,molecule,lee,bracco,rapha,z12,zwid,narpdg}, listed
in Table \ref{Param}. For the heavy quark masses, we could use the range spanned by
the running $\overline{MS}$ mass $\overline{m_Q}(M_Q)$ and the on-shell mass
from QCD (spectral) sum rules compiled in \cite{SNB} and more recently obtained in
Ref.~\cite{SNH10}. However, we do not obtain a valid borel window with the usual 
on-shell mass for $b$ quark, $m_b = 4.70 \GeV$. For this reason, we have considered 
as the maximum value for $b$ quark mass $m_b = 4.60 \GeV$, as indicated in 
Table \ref{Param}. For the $\GGG$ condensate, we have used the new numerical value 
estimated in Ref. \cite{SNH10}.
To take into account the violation of the factorization hypothesis we introduced in 
Table \ref{Param} the parameter $\rho$.

{\small
\begin{table}[t]
\setlength{\tabcolsep}{1.25pc}
\caption{QCD input parameters.}
\begin{tabular}{ll}
&\\
\hline
Parameters&Values\\
\hline
$m_b$ & $(4.24 - 4.60) \GeV$ \\
$m_c$ & $(1.23 - 1.47) \GeV$ \\
$\qq$ & $-(0.23 \pm 0.03)^3\GeV^3$\\
$\lag g_s^2 G^2 \rag$ & $(0.88 \pm 0.25) ~\GeV^4$\\
$m_0^2 \equiv \qGq / \qq$ & $(0.8 \pm 0.1) \GeV^2$\\
$\lag g_s^3 G^3 \rag$ & $(0.58 \pm 0.18)~\GeV^6$\\
$\rho \equiv \qqqq / \qq^2$ & $(0.5 - 2.0)$ \\
\hline
\end{tabular}
\label{Param}
\end{table}}

\begin{figure}[tp]
{\begin{flushleft} a) \end{flushleft}} \vspace{-0.3cm}
\includegraphics[width=7.0cm]{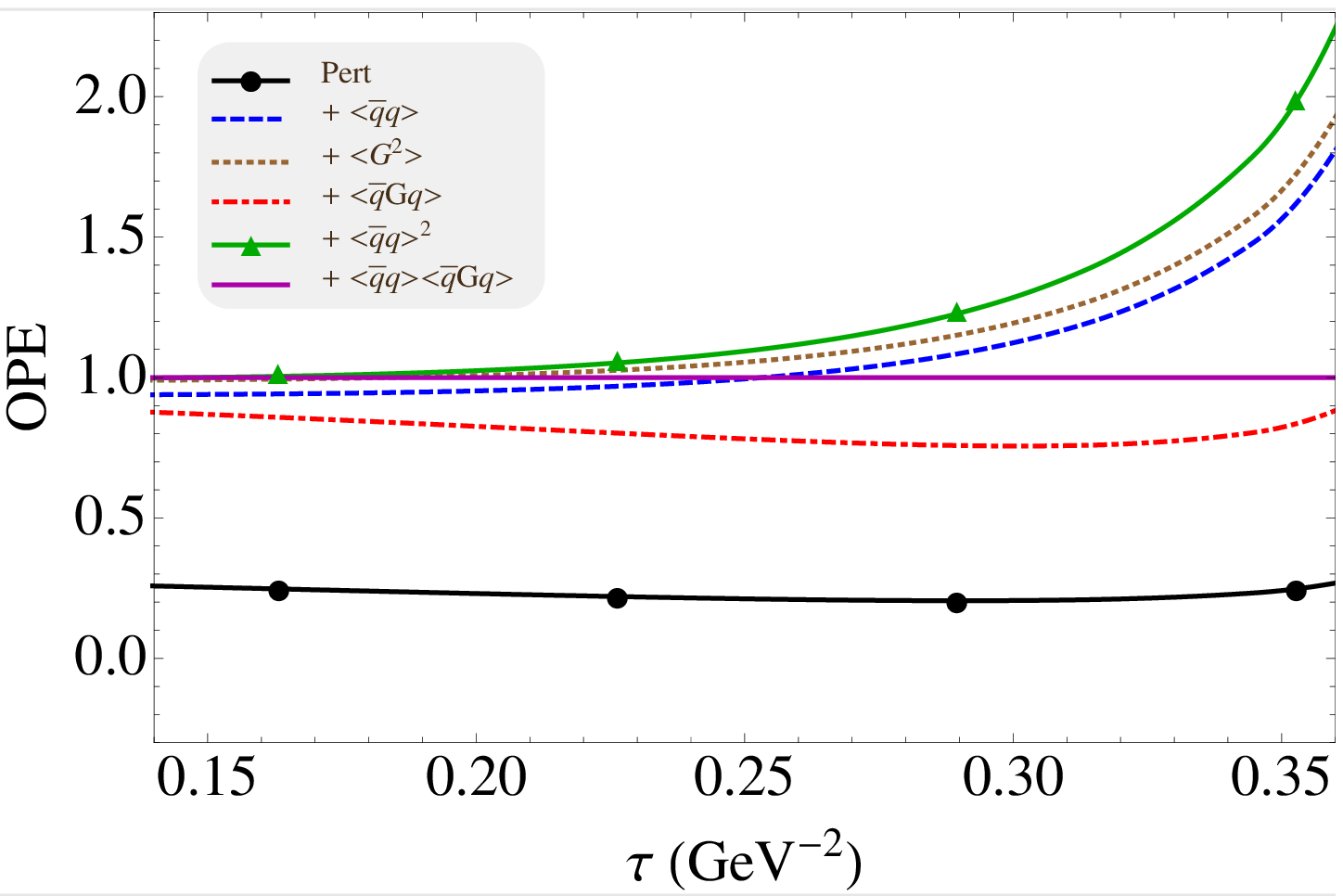}
{\begin{flushleft} b) \end{flushleft}} \vspace{-0.3cm}
\includegraphics[width=7.0cm]{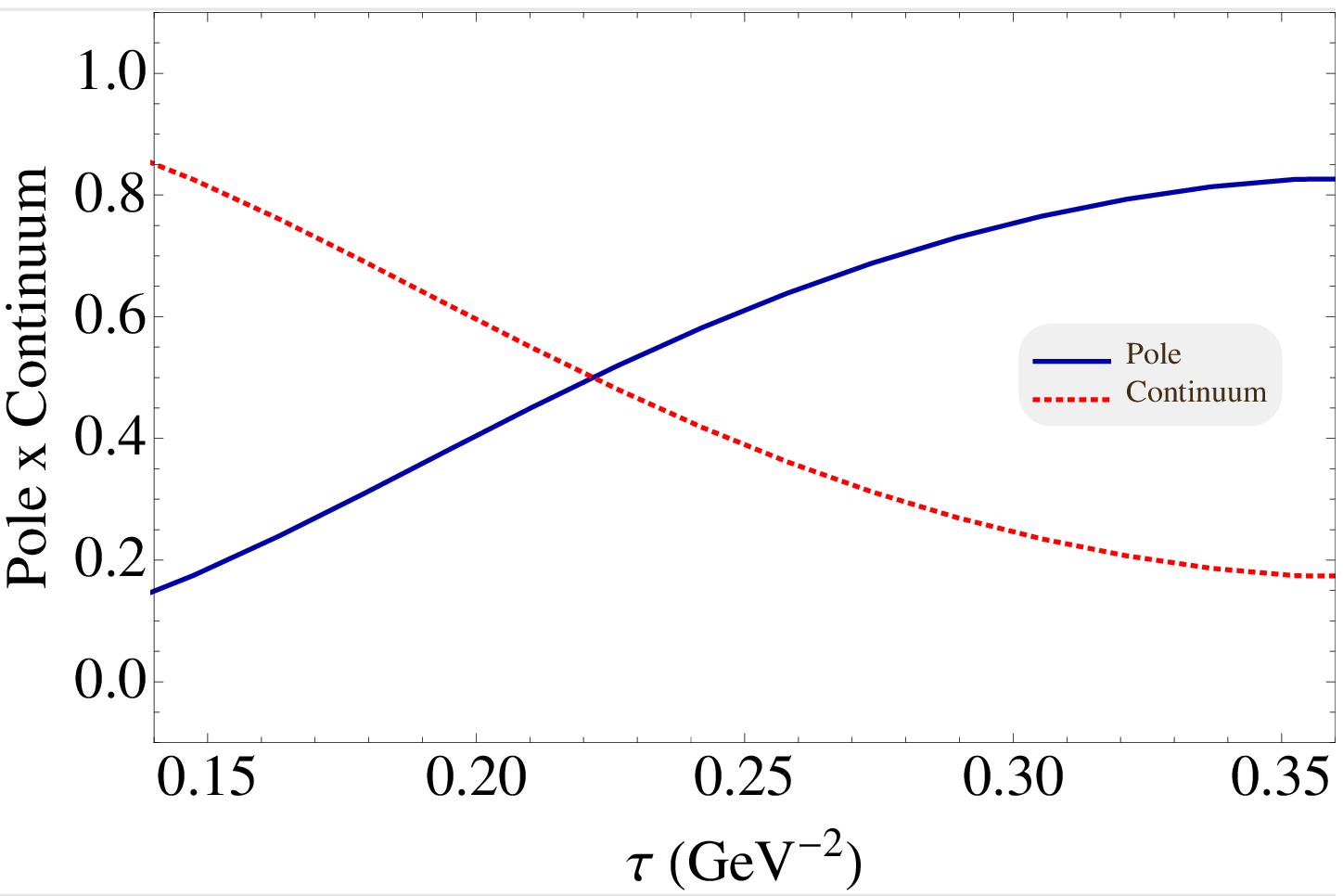}
{\begin{flushleft} c) \end{flushleft}} \vspace{-0.3cm}
\includegraphics[width=7.0cm]{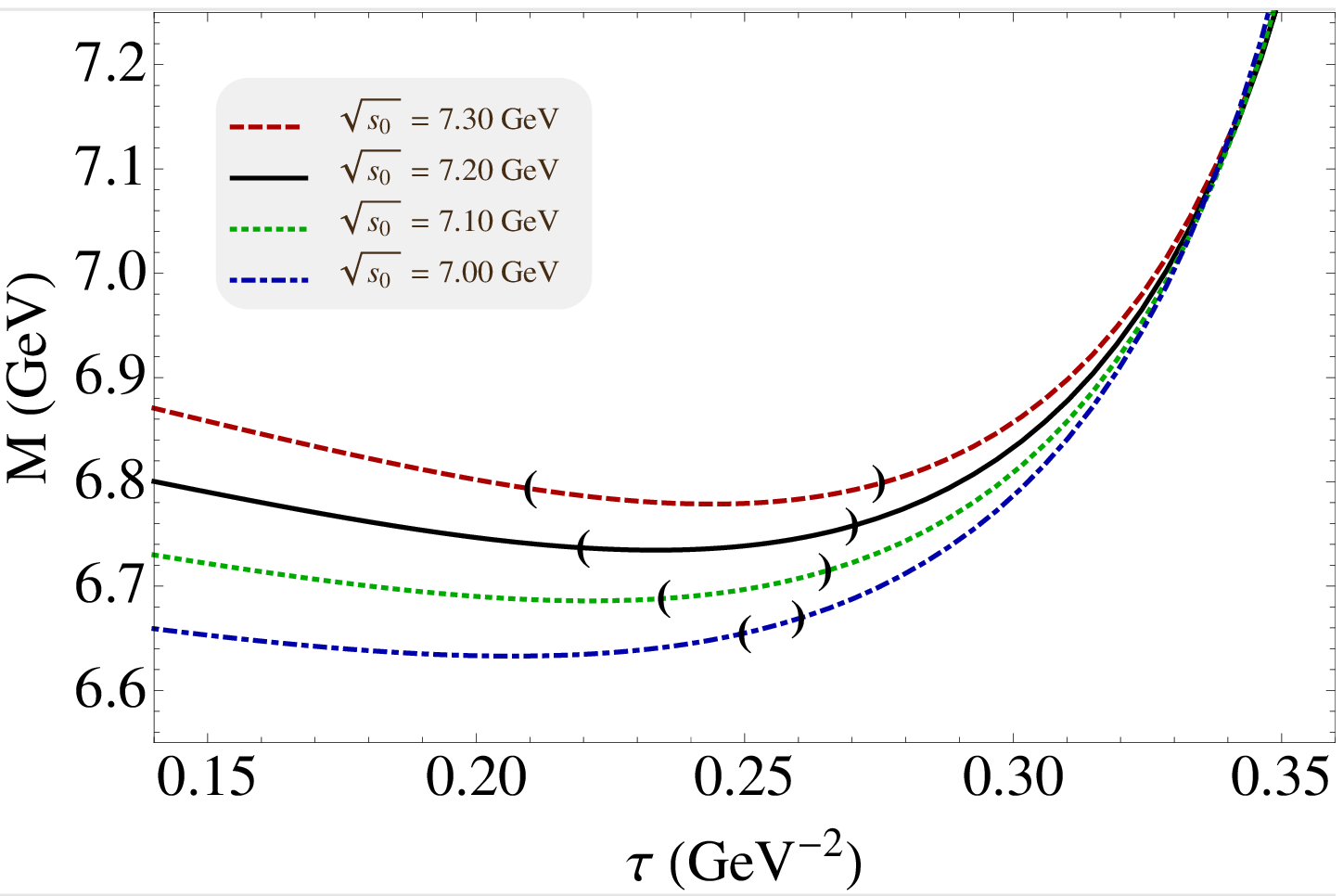}
%
%
\caption{\footnotesize $DB \:(0^+)$ molecular state up to dimension 8 contribution
for $m_c=1.23\GeV$ and $m_b=4.24\GeV$.
{\bf a)} OPE convergence in the region $0.14 \leq \tau \leq 0.36~\GeV^{-2}$ for
$\sqrt{s_0} = 7.20 \GeV$.  
We plot the relative contributions starting with the 
perturbative contribution and each other line represents the relative contribution 
after adding of one extra condensate in the expansion:
+ $\qq$, + $\langle G^2 \rangle$, + $\qGq$, + $\qq^2$ and 
+ $\lag \bar{q}q \rag \lag \bar{q}Gq \rag$.
{\bf b)} The pole and continuum contributions for $\sqrt{s_0} = 7.20 \GeV$.
{\bf c)} The mass as a function of the sum rule parameter $\tau$, 
for different values of $\sqrt{s_0}$.
For each line, the region bounded by parenthesis indicates a valid Borel window.}
\label{DB}
\end{figure}

\begin{figure}[tp]
\begin{center}
{\begin{flushleft} a) \end{flushleft}} 
\includegraphics[width=7.0cm]{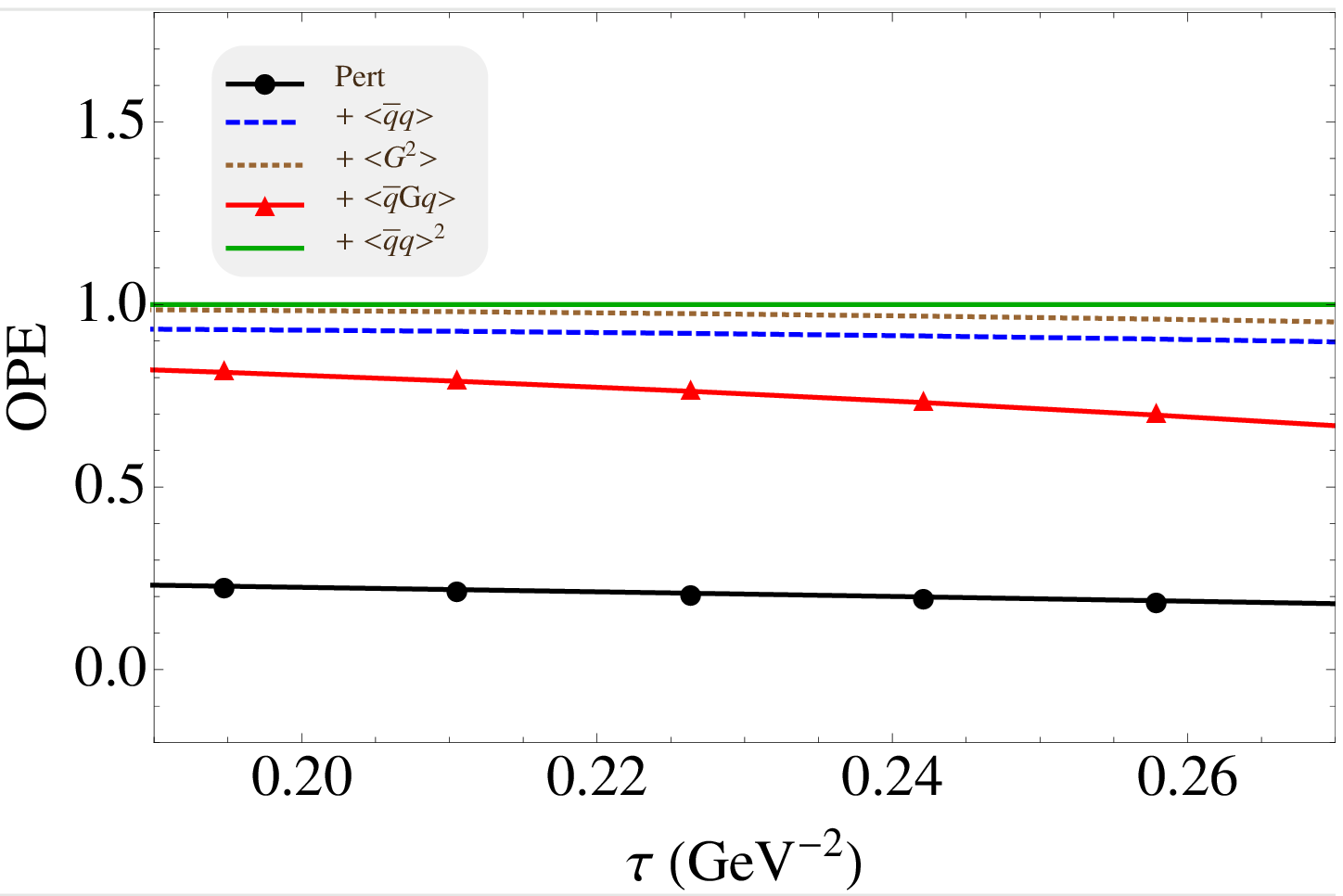}
{\begin{flushleft} b) \end{flushleft}} \vspace{-0.3cm}
\includegraphics[width=7.0cm]{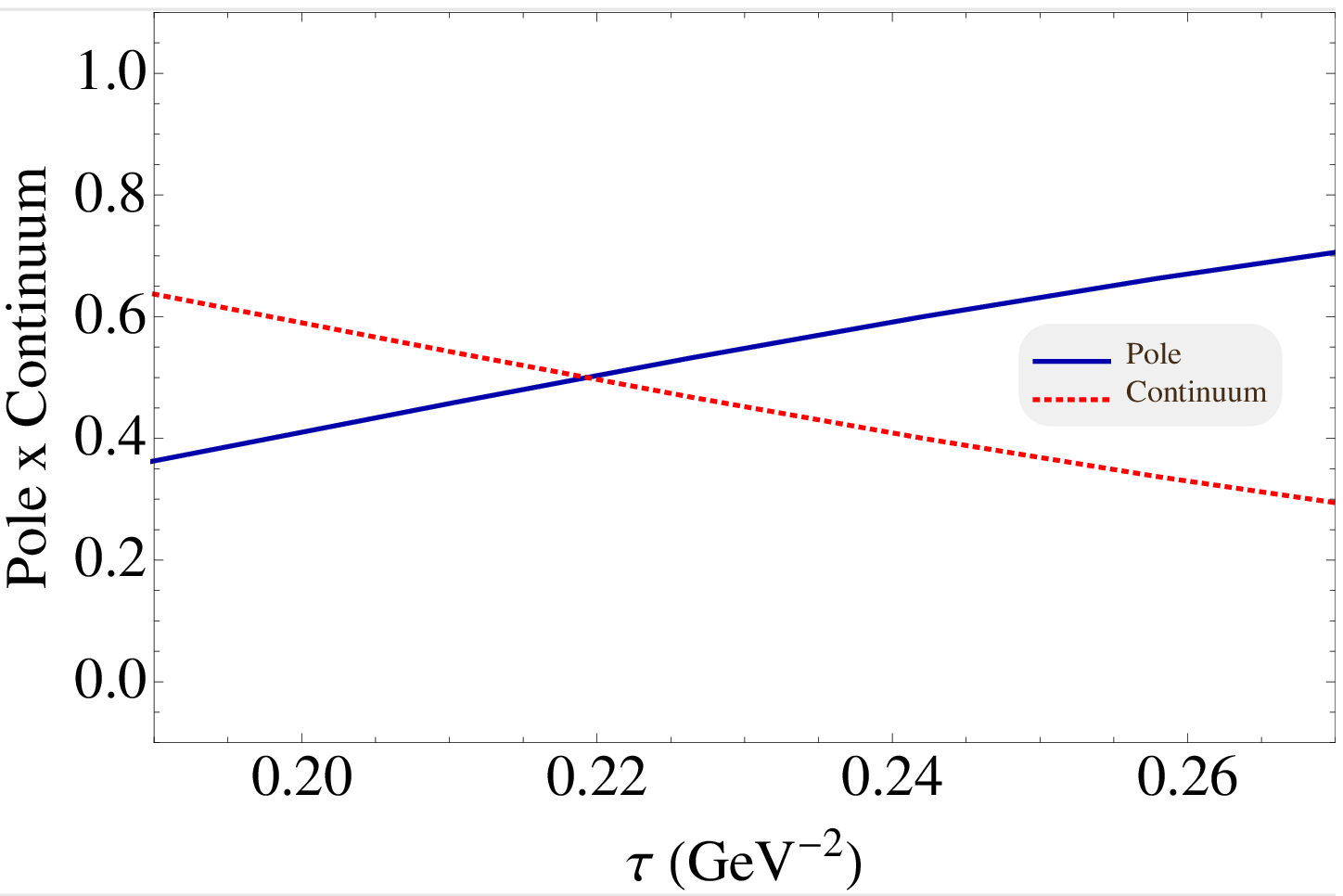}
{\begin{flushleft} c) \end{flushleft}} \vspace{-0.3cm}
\includegraphics[width=7.0cm]{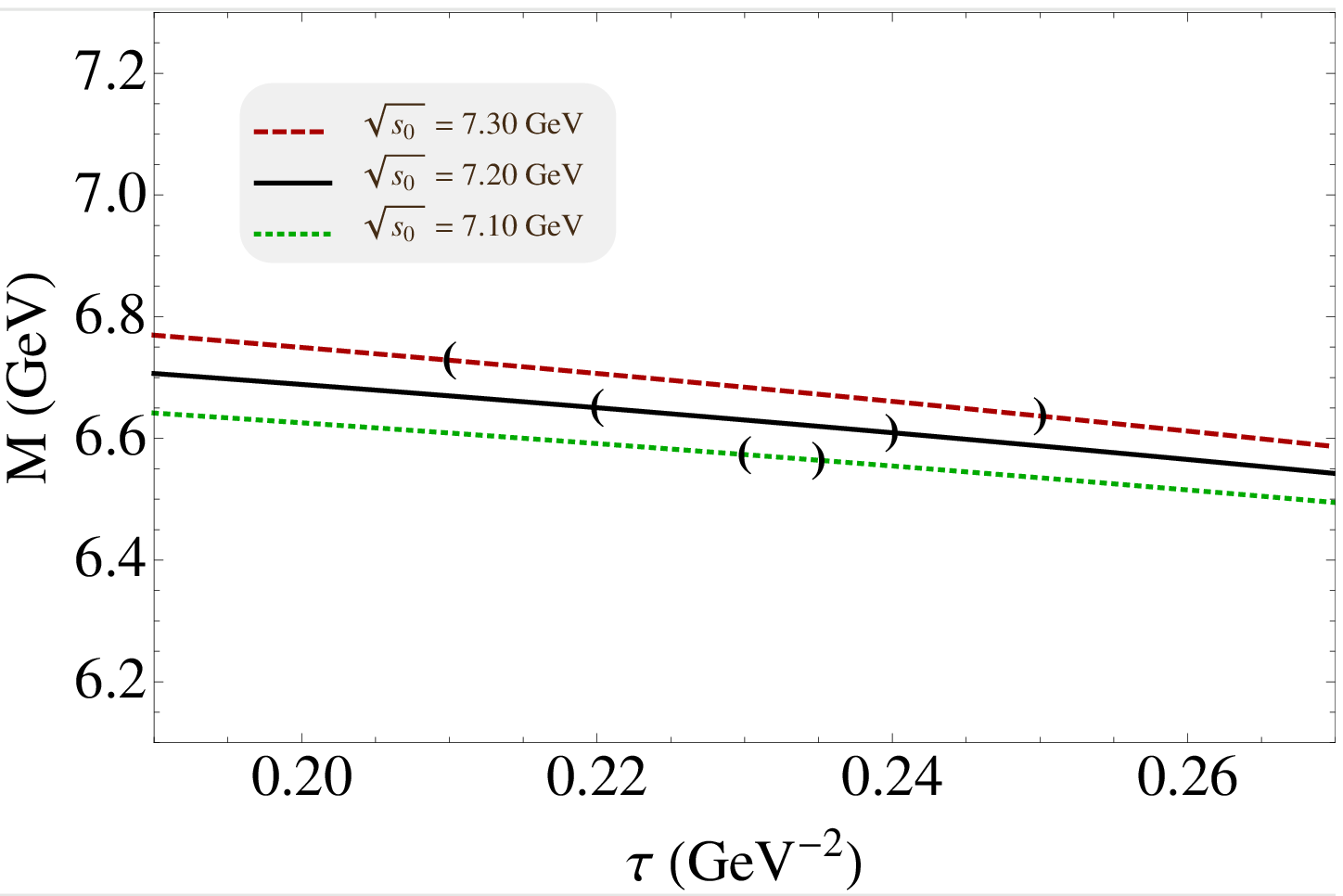}
\caption{\footnotesize  $DB \:(0^+)$ molecular state up to dimension 6 contribution
for $m_c=1.23\GeV$ and $m_b=4.24\GeV$.
{\bf a)} OPE convergence in the region $0.19 \leq \tau \leq 0.27~\GeV^{-2}$ for
$\sqrt{s_0} = 7.20 \GeV$.  
We plot the relative contributions starting with the perturbative contribution and 
each other line represents the relative contribution after adding of one extra 
condensate in the expansion: + $\qq$, + $\langle G^2 \rangle$, + $\qGq$ and 
+ $\qq^2$.
{\bf b)} The pole and continuum contributions for $\sqrt{s_0} = 7.20 \GeV$.
{\bf c)} The mass as a function of the sum rule parameter $\tau$, for 
different values of $\sqrt{s_0}$.
For each line, the region bounded by parenthesis indicates a valid Borel window.}
\label{DB_6}
\end{center}
\end{figure}

Let us consider first the molecular current for the $DB \:(0^+)$  state.
In Fig.~\ref{DB} a),
we show the relative contribution of the terms in the OPE side of the sum rule,
for $\sqrt{s_0} = 7.20 ~\GeV$. From this figure we see that the contribution of the
dimension-8 condensate is smaller than $15 \%$ of the total contribution for
values of $\tau \leq 0.27 \GeV^{-2}$, which indicates a good OPE convergence.
From Fig.~\ref{DB} b), we also see that the pole contribution is bigger than the
continuum contribution only for $\tau \geq 0.22 \GeV^{-2}$. Therefore, we fix
the Borel Window as: $(0.22 \leq \tau \leq 0.27) \GeV^{-2}$.
The results for the mass are shown in Fig.~\ref{DB} c), as a function of $\tau$, 
for different values of $s_0$. 
As we can see from Fig.~\ref{DB} c), the Borel window (indicated
through the parenthesis) gets smaller as the value of $\sqrt{s_0}$ decreases. So, 
we can only work with values for $\sqrt{s_0}$ bigger than $7.00 \GeV$, otherwise 
we do not obtain a valid Borel window for this sum rule.
We also observe that the optimal choice for the continuum threshold is 
$\sqrt{s_0} = 7.20 \GeV$, because it provides the best $\tau-$stability inside of the Borel 
window, including the existence of a minimum point for the value of the mass.

Therefore, varying the value of the continuum threshold in the range 
$\sqrt{s_0}=(7.00 - 7.30)\GeV$, and the others parameters as indicated in Table \ref{Param}, 
we get:
\beqa
  M^{\lag8\rag}_{DB} &=& (6.77 \pm 0.11) \GeV.
  \label{mDB8}
\enqa
The quoted uncertainty is the OPE uncertainty. The most important source of uncertainty
is the values of the heavy quark masses.
As discussed in Ref.~\cite{Lucha:2011zp},
there is another kind of uncertainty, called systematic uncertainty, related to the 
intrinsic limited accuracy of the method. The  systematic uncertainty of the physical 
quantity extracted from the QCDSR represents, perhaps, the most subtle point in the 
application of the method. Without an estimate of the systematic uncertainty,
the numerical value of the physical quantity one reads off from the Borel window might 
differ significantly from its true value. 
In Ref.~\cite{Lucha:2011zp} it was shown that the use of the Borel 
dependent continuum threshold allows to estimate the  systematic uncertainty. In 
particular, for the case of the $D$ and $D_s$ mesons studied in \cite{Lucha:2011zp},
the systematic uncertainty turns out to be of the same order of the OPE uncertainty. 
Since here we do not have how to estimate the Borel dependent continuum threshold,
in an attempt to obtain some information about the systematic uncertainty, 
we will repeat the analysis considering only terms up to dimension 6 in
the OPE. These new results are shown in the Fig.~ \ref{DB_6}.

As one can see in Fig.~\ref{DB_6} a), when we remove the dimension 8 condensates 
contribution we lose the OPE convergence, since the most important contributions to 
the OPE come from $\qq$ and $\rho \qq^2$ contributions. Thus to be able to extract 
some results from this analysis we determine the maximum value of $\tau$ parameter 
imposing that the contribution of the dimension 6 condensate is smaller than 25\% 
of the total contribution, otherwise we do not have a valid Borel window
for this sum rule. The minimum value of $\tau$ is not changed since the pole 
dominance behavior remains the same. Finally, we obtain the results shown in the 
Fig.~ \ref{DB_6} c), from where we get:
\beqa
  M^{\lag6\rag}_{DB} &=& (6.63 \pm 0.09) \GeV.
  \label{mDB6}
\enqa

Note that the value in Eq.(\ref{mDB6}) differs at maximum only $\sim$5.0\% to 
that in Eq.(\ref{mDB8}). Besides, the 
inclusion of dimension 8 condensate provides a better OPE convergence, 
$\tau$-stability and an improved Borel window. Therefore, even being aware that
this is only part of the dimension-8 contribution, here we consider it as a form
to estimate the systematic uncertainty. One should note that a complete
evaluation of the dimension-8 contributions require more involved analysis
including a non-trivial choice of the factorization assumption basis \cite{BAGAN}.
Then, the final value for the $DB$ molecular state is given by:
\beqa
  M_{DB} &=& (6.75 \pm 0.14) \GeV.
  \label{mDB}
\enqa

The mass in Eq.~(\ref{mDB}) is $\sim \!\!400 \MeV$ below the $DB$ threshold
indicating that such molecular state would be tightly bound. This result, for the
binding energy, is very different than the obtained in Ref.~\cite{Sun:2012sy} for the
$DB \:(0^+)$  molecular state. The authors of  Ref.~\cite{Sun:2012sy} found that
the $DB \:(0^+)$  molecular state is loosely bound with a binding energy
smaller than $14~\MeV$. However, it is very important to notice that since
the molecular currents given in Table \ref{mol} are local, they do
not represent extended objects, with two mesons separated in space, but
rather a very compact object with two singlet quark-antiquark pairs. Therefore,
the result obtained here may suggest that, although a loosely bound
$DB \:(0^+)$  molecular state can exist, it may not be the ground state
for a four-quark exotic state with the same quantum numbers and quark content.

Having the hadron mass, we can also evaluate the coupling parameter, $\lambda$,
defined in the Eq.(\ref{o+}). We get:
\beqa
  \lambda_{DB} &=& (0.029 \pm 0.008) \GeV^5 ~~.
  \label{lDB}
\enqa
The  parameter $\lambda$ gives a measure of the
strength of the coupling between the current and the state. 
The result in Eq.~(\ref{lDB}) has the same order of magnitude as the coupling obtained
for the $X(3872)$ \cite{x3872}, for example. This indicates that such state could 
be very well represented by the respective current in Table \ref{mol}.

\begin{figure}[tp]
\begin{center}
{\begin{flushleft} a) \end{flushleft}} \vspace{-0.3cm}
\includegraphics[width=7.0cm]{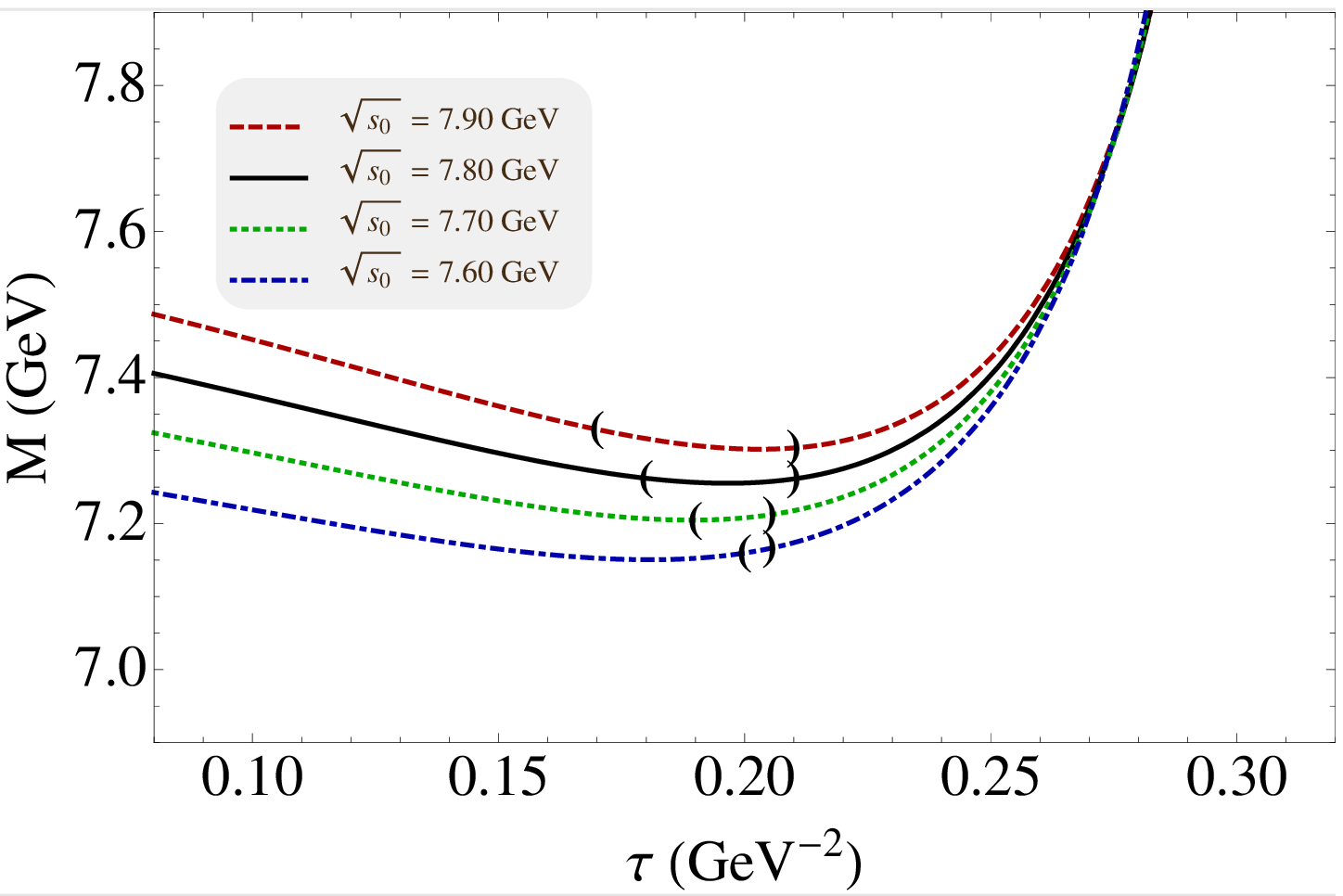}
{\begin{flushleft} b) \end{flushleft}} \vspace{-0.3cm}
\includegraphics[width=7.0cm]{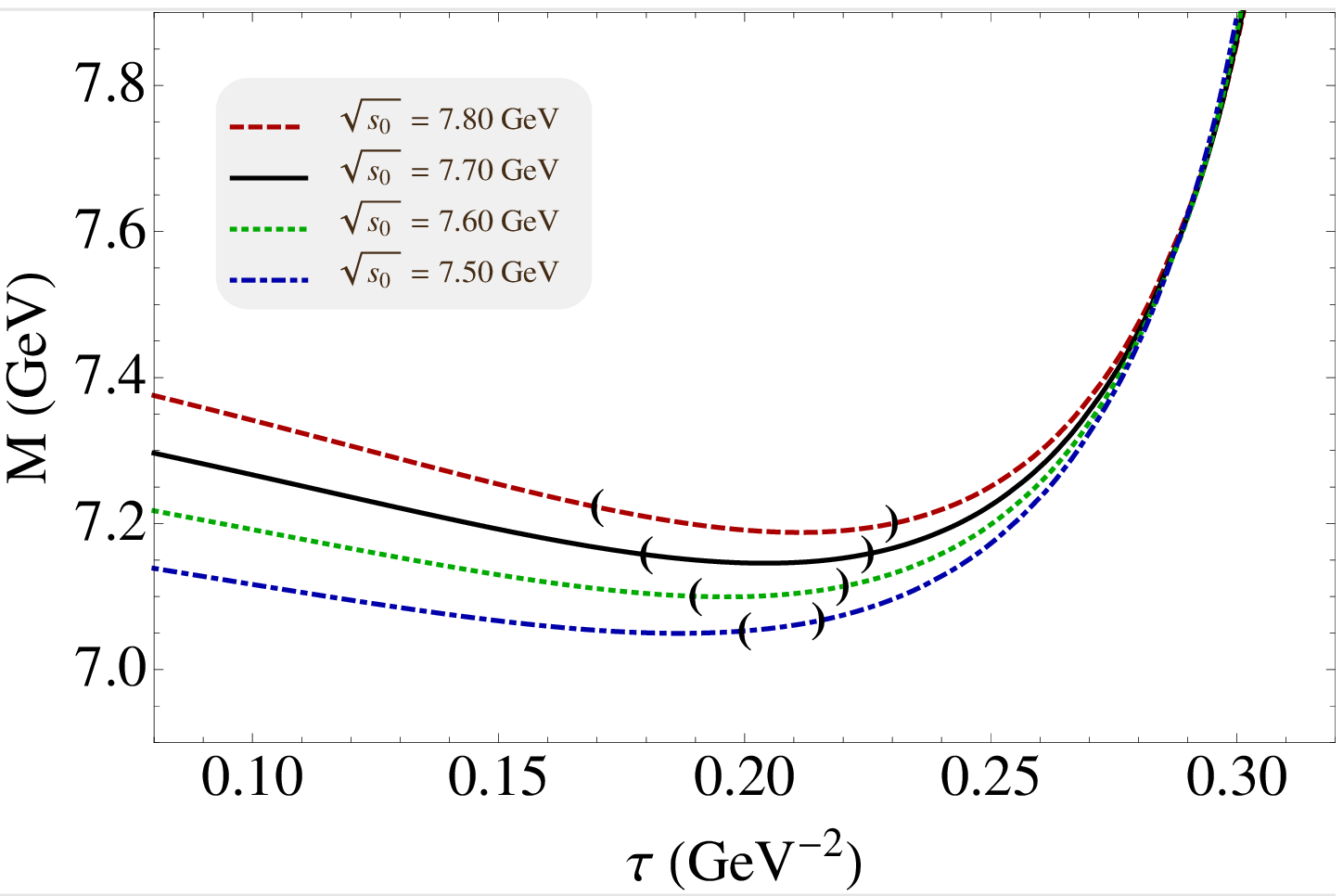}
{\begin{flushleft} c) \end{flushleft}} \vspace{-0.3cm}
\includegraphics[width=7.0cm]{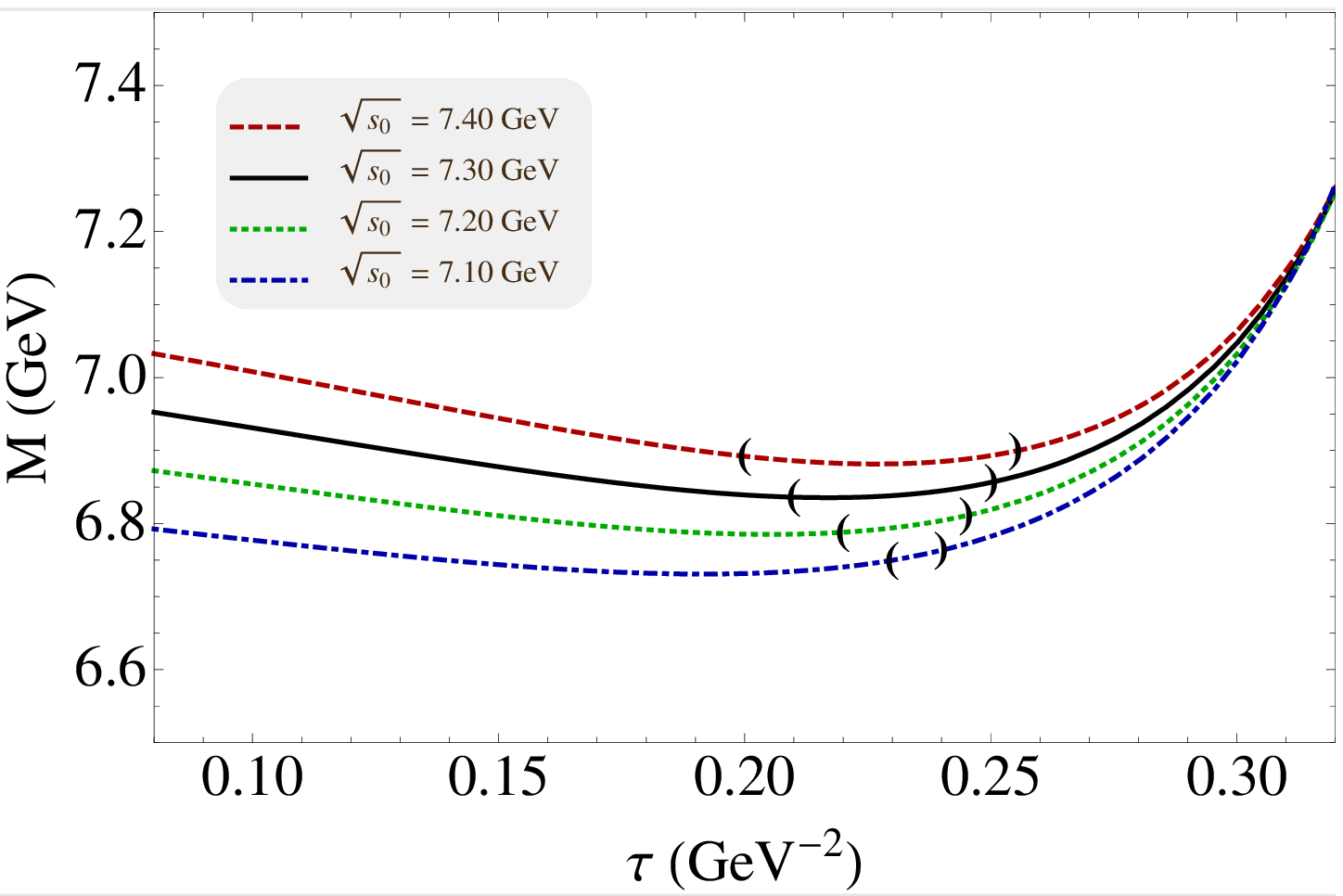}
\caption{\footnotesize  The mass as a function of the sum rule
parameter $\tau$, for $m_c=1.23\GeV$ and $m_b=4.24\GeV$, 
considering different values for $\sqrt{s_0}$:
{\bf a)} for $D^*B^*$, $0^+$ molecular current;
{\bf b)} for $D^*B$, $1^+$ molecular current;
{\bf c)} for $DB^*$, $1^+$ molecular current.
For each line, the region bounded by parenthesis indicates a valid Borel window.}
\label{DxBx}
\end{center}
\end{figure}

We can extend the same analysis to study the others molecular states presented in
Table \ref{mol}. For all of them we get a similar OPE convergence in a region
where the pole contribution is bigger than the continuum contribution.
We obtain the results shown in the Fig.~ \ref{DxBx}.

In Fig.~\ref{DxBx} a), we show the ground state mass, for the $D^*B^*$, $0^+$ molecular
current, as a function of $\tau$.
For $\sqrt{s_0} = 7.80~\GeV$, we can fix the Borel window
as: $(0.18 \leq \tau \leq 0.21) \GeV^{-2}$.
>From this figure we again see that there is a very good $\tau$-stability in the
determined Borel window.

Varying the value of the continuum threshold in the range
$\sqrt{s_0} = (7.60 - 7.90) \GeV$,  the others parameters as indicated in
Table \ref{Param} and also estimating the uncertainty by neglecting the
dimension-8 contribution  we get:
\beqa
 M_{D^*B^*} &=& (7.27\pm 0.12)~\GeV, \\
 \lambda_{D^*B^*} &=& (0.115\pm 0.021)~\GeV^5  ~~.
 \label{mDxBx}
\enqa
The obtained mass indicates a binding energy of the order of
$\sim \! 50\MeV$ below the $D^*B^*$ threshold. Considering the uncertainties, 
it is even possible that this state is not bound. In this case, our central 
result is in a good agreement with the result obtained
for the $D^*B^*$, $0^+$, molecular state obtained in Ref.~\cite{Sun:2012sy}. 
However, since we do not have a trustable estimate for the systematic error, 
as discussed above, any conclusion about the possible existence of this
state would be premature.

We now consider the $D^*B \:(1^+)$ molecular current.
In Fig.~\ref{DxBx} b), we  show the ground state mass, as a function of $\tau$.
For $\sqrt{s_0} = 7.70~\GeV$, we can fix the Borel window
as: $(0.18 \leq \tau \leq 0.22) \GeV^{-2}$.

Varying the value of the continuum threshold in the range $\sqrt{s_0} = (7.50 - 7.80) \GeV$,  
the others parameters as indicated in
Table \ref{Param} and also estimating the uncertainty by neglecting the
dimension-8 contribution  we get:
\beqa
 M_{D^*B} &=& (7.16\pm0.12)~\GeV, \\
 \lambda_{D^*B} &=& (0.058\pm 0.013)~\GeV^5  ~~.
 \label{mDxB}
\enqa
The obtained mass indicates a central binding energy for the  $D^*B$, $1^+$ state of
the order of $\sim \! 130 \MeV$.
Considering the uncertainty, this result might be compatible with the one obtained by the
authors in Ref. \cite{Sun:2012sy}, or can even be unbound, as the $D^*B^*$ state.
Therefore, also in this case, any conclusion about the possible existence of this
state would be premature.

Finally, we study the molecular current for $DB^*$, $1^+$ state. As one can see from
 Fig.~ \ref{DxBx} c), we have a very good
$\tau$-stability inside of the Borel window: $(0.21 \leq \tau \leq 0.25) \GeV^{-2}$,
for $\sqrt{s_0} = 7.30~\GeV$. Doing the same procedure to estimate the
uncertainties in the range
$\sqrt{s_0} = (7.10 - 7.40) \GeV$ we get:
\beqa
 M_{DB^*} &=& (6.85\pm0.15)~\GeV,
 \label{mDBx}\\
 \lambda_{DB^*} &=& (0.036\pm 0.011)~\GeV^5
 \label{lDBx}
\enqa
which indicates a binding energy of the order $\sim \! 330\MeV$, much bigger than 
that obtained in Ref.~\cite{Sun:2012sy}. 

We can compare our results with the ones presented in Ref.~\cite{Zhang}.
First of all we would like to point out that we have found some disagreements 
in the spectral densities expressions for the $DB$, $D^* B$ and $D B^*$ molecular 
currents. In particular, we have found some missing terms in the $\GG$, $\qGq$ 
and $\GGG$ contributions, due to some diagrams  that have been neglected in 
their calculations. We have found that the $\qGq$ contribution plays an important 
role to the final result, and this can explain why the mass values found in 
Ref.\cite{Zhang} differ from ours. Another important point, in which our calculations
differ, is the fact that the Borel window $(0.10 \leq \tau \leq 0.14) \GeV^{-2}$, 
considered by the authors in Ref.\cite{Zhang}, does not have pole dominance, as 
can be seen in Figs. \ref{DB} c), \ref{DxBx} a), \ref{DxBx} b) and \ref{DxBx} c). 
The only result for the mass, which is in agreement with Ref.\cite{Zhang}, 
is the one  for the $D^*B^*$ molecular current. For this current we 
found disagreements only for 
the $\GGG$ contribution. Since the  $\GGG$ contribution is very small, as compared 
to the others, the differences found could not modify the final result.

In conclusion, we have studied the mass of the exotic $B_c$-like molecular states
using QCD sum rules. We find that for the molecular currents $D^*B^* \:(J^P=0^+)$ and
$D^*B \:(J^P=1^+)$, the QCDSR central results lead approximately to the same 
predictions made by the authors in Ref.~\cite{Sun:2012sy}, for the respective 
molecular states in a OBE model. However, since our uncertainties are underestimated
due to our crude model for the continuum threshold, any conclusion about the possible 
existence of these states would be premature.

In the case of the  $DB \:(J^P=0^+)$ and $DB^* \:(J^P=1^+)$ molecular
currents, from the QCD sum rule point of view, the masses of the corresponding states
are smaller than the masses obtained for the respective molecular states studied in
Ref.~\cite{Sun:2012sy}. We interpret this result as an indication of the possible
existence of four-quark states, with the same quark content and quantum numbers as
the $DB \:(J^P=0^+)$ and $DB^* \:(J^P=1^+)$ molecular states, but with smaller masses.

\section*{Acknowledgment}
This project is supported by the National Natural Science Foundation of
China under Grants 11175073, 11035006, the
Ministry of Education of China (FANEDD under Grant No. 200924,
DPFIHE under Grant No. 20090211120029, NCET, the Fundamental
Research Funds for the Central Universities), the Fok Ying-Tong Education Foundation 
(No. 131006) and CNPq and FAPESP-Brazil.

\appendix


\section{Spectral Densities}\label{App1}

The spectral densities expressions for the molecular currents given in
Table \ref{mol}, were calculated up to dimension-6 condensates, at leading
order in $\alpha_s$. To keep the heavy
quark mass finite, we use the momentum-space expression for the heavy quark
propagator. We calculate the light quark part of the correlation
function in the coordinate-space, and we use the Schwinger parameters to
evaluate the heavy quark part of the correlator. To evaluate the $d^4x$
integration in Eq.~(\ref{2po}), we use again the Schwinger parameters, after
a Wick rotation. Finally we get integrals in the Schwinger parameters.
The result of these integrals are given in
terms of logarithmic functions, from where we extract the spectral densities
and the limits of the integration. The same technique can be used to evaluate
the condensate contributions. To evaluate the systematic uncertainty we also include
a part of the dimension-8 contribution, related with the mixed-condensate times
the quark condensate. In Ref.~\cite{Finazzo:2011he} it was shown that the 
contribution of this condensate is much bigger than other dimension-8 
condensates, related with the gluon condensate. 

For the  $DB$, $0^+$ molecular current we get:
\beqa
  \rho^{pert}_{_{DB}}(s) &=& \frac{3}{2^{11} \pi^6}
  	\int\limits^{\almax}_{\almin} \!\!\frac{d\al}{\alpha^3}  
	\int\limits^{\bemax}_{\bemin} \!\!\frac{d\be}{\beta^3} \:
	(1-\alpha -\beta) F(\al,\be)^4 \nn,\\
  \rho^{\qq}_{_{DB}}(s) &=& -\frac{3 \qq}{2^{7} \pi^4}
  	\int\limits^{\almax}_{\almin} \!\!\frac{d\al}{\alpha^2}  \int\limits^{\bemax}_{\bemin} \!\!\frac{d\be}{\beta^2} \:
	 (\be \:m_c + \al \:m_b) F(\al,\be)^2 \nn,\\
  \rho^{\GG}_{_{DB}}(s) &=& \frac{\gG}{2^{12} \pi^6}
  	\int\limits^{\almax}_{\almin} \!\!\frac{d\al}{\alpha^3}  \int\limits^{\bemax}_{\bemin} \!\!\frac{d\be}{\beta^3} \:
	F(\al,\be) \bigg[ 3 \al \be (\alpha +\beta) F(\al,\be) \nn\\
	&& +~ 2 (1-\al-\be) (\be^3 m_c^2 + \al^3 m_b^2 ) \bigg] \nn,\\
  \rho^{\qGq}_{_{DB}}(s) &=& - \frac{3\qGq}{2^{8} \pi^4} \Bigg[
  	\int\limits^{\almax}_{\almin} \!\!\!\!\frac{d\al}{\al(1\!-\! \al)} \bigg( m_c \!+\! \al(m_b \!-\! m_c) \bigg) H(\al) \nn\\
	&& -~ 2 \!\!\int\limits^{\almax}_{\almin} \!\!\frac{d\al}{\al^2} \!\!
	\int\limits^{\bemax}_{\bemin} \!\! \frac{d\be}{\be^2} ( \be^2 m_c \!+\! \al^2 m_b ) F(\al,\be) \Bigg], ~~~~\nn\\
  \rho^{\qq^2}_{_{DB}}(s) &=& \frac{m_b m_c\rho \qq^2}{16 \pi^2} \lambda_{bc} \:v \nn,\\
  \rho^{\GGG}_{_{DB}}(s) &=& \frac{\gGGG}{2^{13} \pi^6}
  	\!\!\int\limits^{\almax}_{\almin} \!\!\frac{d\al}{\al^3} \!\!
	\int\limits^{\bemax}_{\bemin} \!\! \frac{d\be}{\be^3} (1-\al-\be) \bigg[ 
	2 (\be^4 m_c^2 \!+\! \al^4 m_b^2 ) \nn\\
	&& +~ (\al^3 + \be^3)F(\al,\be) ~\bigg], \nn\\
  \rho^{\8}_{_{DB}}(s) &=& \frac{m_c m_b\rho \qq\qGq}{2^{5} \pi^2}
  	\int\limits^{1}_{0} \!\!\frac{d\al}{\al(1-\al)} \Bigg[ 1-\al+\al^2 \nn\\
	&& -~ \bigg( m_c^2 -\al (m_c^2 - m_b^2) \bigg) \tau \Bigg]
	\:\delta\bigg( s - \frac{m_c^2 -\al( m_c^2 - m_b^2)}{\al(1-\al)}\bigg) \nn.
\enqa

For the  $D^*B$, $1^+$ molecular current we get:
\beqa
  \rho^{pert}_{_{D^*B}}(s) &=& \frac{3}{2^{12} \pi^6}
  	\int\limits^{\almax}_{\almin} \!\!\frac{d\al}{\alpha^3}  \!\!\int\limits^{\bemax}_{\bemin} \!\!\frac{d\be}{\beta^3} \:
	(1 \!-\! \alpha \!-\! \beta)(1\!+\! \al \!+\! \be) F(\al,\be)^4, \nn\\
  \rho^{\qq}_{_{D^*B}}(s) &=& -\frac{3 \qq}{2^{7} \pi^4}
  	\!\!\int\limits^{\almax}_{\almin} \!\!\frac{d\al}{\alpha^2}  \!\!\int\limits^{\bemax}_{\bemin} \!\!\frac{d\be}{\beta^2} \:
	 \bigg[ \be \:m_c + \al(\al+\be) \:m_b \bigg] F(\al,\be)^2,\nn\\
  \rho^{\GG}_{_{D^*B}}(s) &=& \frac{\gG}{2^{12} \pi^6}
  	\!\!\int\limits^{\almax}_{\almin} \!\!\frac{d\al}{\alpha^3}  \!\!\int\limits^{\bemax}_{\bemin} \!\!\frac{d\be}{\beta^3} \: F(\al,\be) \Bigg[
	\al \be \bigg( 3\al (\al +\be) \nn\\
	&-& \be(2-\al -\be) \bigg) F(\al,\be)  + ( \be^3 m_c^2 + \al^3 m_b^2 ) \nn\\
	&\times&(1-\al-\be)(1+\al+\be) \Bigg], \nn\\
  \rho^{\qGq}_{_{D^*B}}(s) &=& - \frac{3\qGq}{2^{8} \pi^4} \Bigg[
  	\int\limits^{\almax}_{\almin} \!\!\!\!\frac{d\al}{\al(1\!-\! \al)} \bigg( m_c - \al(m_c-m_b) \bigg) H(\al) \nn\\
	&-& m_b \!\!\int\limits^{\almax}_{\almin} \!\!d\al \!\!
	\int\limits^{\bemax}_{\bemin} \!\!\frac{d\be}{\be^2} \: (2\al+3\be)F(\al,\be) \Bigg], ~~~~\nn\\
  \rho^{\qq^2}_{_{D^*B}}(s) &=& \frac{m_c m_b\rho \qq^2}{16 \pi^2} \lambda_{bc} \:v, \nn\\
  \rho^{\GGG}_{_{D^*B}}(s) &=& \frac{\gGGG}{2^{14} \pi^6}
  	\!\!\int\limits^{\almax}_{\almin} \!\!\frac{d\al}{\al^3} \!\!
	\int\limits^{\bemax}_{\bemin} \!\! \frac{d\be}{\be^3} (1-\al-\be)(1+\al+\be) \bigg[ \nn\\
	&& \times~ 2 (\be^4 m_c^2 \!+\! \al^4 m_b^2 ) + (\al^3 + \be^3)F(\al,\be) ~\bigg], \nn\\
  \rho^{\8}_{_{D^*B}}(s) &=& \frac{m_c m_b\rho \qq\qGq}{2^{5} \pi^2}
  	\int\limits^{1}_{0} \!\!\frac{d\al}{\al(1 \!-\! \al)} \delta\bigg( s - \frac{m_c^2 -\al( m_c^2 \!-\! m_b^2)}{\al(1-\al)}\bigg) \nn\\
	&\times&\Bigg[ \al^2 - \bigg( m_c^2 -\al (m_c^2 - m_b^2) \bigg) \tau \Bigg]. \nn
\enqa

For the  $DB^*$, $1^+$ molecular current we get:
\beqa
  \rho^{pert}_{_{DB^*}}(s) &=& \frac{3}{2^{12} \pi^6}
  	\int\limits^{\almax}_{\almin} \!\!\frac{d\al}{\alpha^3}  \int\limits^{\bemax}_{\bemin} \!\!\frac{d\be}{\beta^3} \:
	(1 \!-\! \alpha \!-\! \beta)(1\!+\! \al \!+\! \be) F(\al,\be)^4, \nn\\
  \rho^{\qq}_{_{DB^*}}(s) &=& -\frac{3 \qq}{2^{7} \pi^4}
  	\int\limits^{\almax}_{\almin} \!\!\frac{d\al}{\alpha^2}  \int\limits^{\bemax}_{\bemin} \!\!\frac{d\be}{\beta^2} \:
	 \bigg[ \be(\al+\be) \:m_c + \al \:m_b \bigg] F(\al,\be)^2, \nn\\
  \rho^{\GG}_{_{DB^*}}(s) &=& \frac{\gG}{2^{12} \pi^6}
  	\int\limits^{\almax}_{\almin} \!\!\frac{d\al}{\alpha^3}  \int\limits^{\bemax}_{\bemin} \!\!\frac{d\be}{\beta^3} \: F(\al,\be) \Bigg[
	\al \be \bigg( 3\be (\al +\be) \nn\\
	&-& \al(2-\al -\be) \bigg) F(\al,\be) + ( \be^3 m_c^2 + \al^3 m_b^2 ) \nn\\
	&\times&(1-\al-\be)(1+\al+\be) \Bigg], \nn\\
  \rho^{\qGq}_{_{DB^*}}(s) &=& - \frac{3\qGq}{2^{8} \pi^4} \Bigg[
  	\int\limits^{\almax}_{\almin} \!\!\frac{d\al}{\al(1 \!-\! \al)} \bigg( m_c - \al(m_c - m_b) \bigg) H(\al)\nn\\
	&-& m_c \!\!\int\limits^{\almax}_{\almin} \!\!\frac{d\al}{\al^2} \!\!
	\int\limits^{\bemax}_{\bemin} \!\! d\be \: (3\al+2\be)F(\al,\be) \Bigg], ~~~~\nn\\
  \rho^{\qq^2}_{_{DB^*}}(s) &=& \frac{m_c m_b\rho \qq^2}{16 \pi^2} \lambda_{bc} \:v, \nn\\
  \rho^{\GGG}_{_{DB^*}}(s) &=& \frac{\gGGG}{2^{14} \pi^6}
  	\!\!\int\limits^{\almax}_{\almin} \!\!\frac{d\al}{\al^3} \!\!
	\int\limits^{\bemax}_{\bemin} \!\! \frac{d\be}{\be^3} (1-\al-\be)(1+\al+\be) \bigg[ \nn\\
	&& \times~ 2 (\be^4 m_c^2 \!+\! \al^4 m_b^2 ) + (\al^3 + \be^3)F(\al,\be) ~\bigg], \nn\\
  \rho^{\8}_{_{DB^*}}(s) &=& \frac{m_c m_b\rho \qq\qGq}{2^{5} \pi^2} \!\!\!
  	\int\limits^{1}_{0} \!\!\!\!\frac{d\al}{\al(1 \!-\! \al)} \delta\bigg( s - \frac{m_c^2 \!-\! \al( m_c^2 \!-\! m_b^2)}{\al(1-\al)}\bigg) \nn\\
	&\times&\Bigg[ (1-\al)^2 - \bigg( m_c^2 -\al (m_c^2 - m_b^2) \bigg) \tau \Bigg]. \nn
\enqa

For the  $D^*B^*$, $0^+$ molecular current we get:
\beqa
  \rho^{pert}_{_{D^*B^*}}(s) &=& \frac{3}{2^{9} \pi^6}
  	\int\limits^{\almax}_{\almin} \!\!\frac{d\al}{\alpha^3}  \int\limits^{\bemax}_{\bemin} \!\!\frac{d\be}{\beta^3} \:
	(1-\alpha -\beta) F(\al,\be)^4, \nn\\
  \rho^{\qq}_{_{D^*B^*}}(s) &=& -\frac{3 \qq}{2^{6} \pi^4}
  	\int\limits^{\almax}_{\almin} \!\!\frac{d\al}{\alpha^2}  \int\limits^{\bemax}_{\bemin} \!\!\frac{d\be}{\beta^2} \:
	(\be \:m_c + \al \:m_b) F(\al,\be)^2, \nn\\
  \rho^{\GG}_{_{D^*B^*}}(s) &=& \frac{\gG}{2^{9} \pi^6}
  	\int\limits^{\almax}_{\almin} \!\!\frac{d\al}{\alpha^3}  \int\limits^{\bemax}_{\bemin} \!\!\frac{d\be}{\beta^3} \:
	(1-\al-\be) \nn\\
	&& \times~ ( \be^3 m_c^2 + \al^3 m_b^2 ) F(\al,\be), \nn\\
  \rho^{\qGq}_{_{D^*B^*}}(s) &=& - \frac{3 \qGq}{2^{7} \pi^4}
  	\int\limits^{\almax}_{\almin} \!\!\frac{d\al}{\al(1-\al)} \:\bigg( m_c - \al(m_c-m_b) \bigg) H(\al), \nn\\
  \rho^{\qq^2}_{_{D^*B^*}}(s) &=& \frac{m_c m_b\rho \qq^2}{4 \pi^2} \lambda_{bc} \:v, \nn\\
  \rho^{\GGG}_{_{D^*B^*}}(s) &=& \frac{\gGGG}{2^{11} \pi^6}
  	\!\!\int\limits^{\almax}_{\almin} \!\!\frac{d\al}{\al^3} \!\!
	\int\limits^{\bemax}_{\bemin} \!\! \frac{d\be}{\be^3} (1-\al-\be) \bigg[ 
	2 (\be^4 m_c^2 \!+\! \al^4 m_b^2 )\nn\\
	&& +~ (\al^3 + \be^3)F(\al,\be) ~\bigg], \nn\\
  \rho^{\8}_{_{D^*B^*}}(s) &=& - \frac{m_c m_b\rho \qq\qGq}{8 \pi^2} \!\!\!
  	\int\limits^{1}_{0} \!\!\!\!\frac{d\al}{\al(1 \!-\! \al)} \delta\bigg( s - \frac{m_c^2 \!-\! \al( m_c^2 \!-\! m_b^2)}{\al(1-\al)}\bigg) \nn\\
	&\times&\Bigg[ \al(1-\al) + \bigg( m_c^2 -\al (m_c^2 - m_b^2) \bigg) \tau \Bigg]. \nn
\enqa

In all these expressions we have used the following definitions:
\beqa
  H(\al) &=& m_b^2 \al + m_c^2(1-\al) - \al(1-\al) s, \\ && \nn\\
  F(\al,\be) &=& m_b^2 \al + m_c^2 \be - \al\be s, \\ && \nn\\
  \lambda_{bc} &=& 1 + (m_c^2 - m_b^2) / s,\\ && \nn\\
  v &=& \sqrt{1 - \frac{4m_c^2 / s}{\lambda_{bc}^2}},
\enqa
and the integration limits are given by:
\beqa
  \be_{min} &=& \frac{\al \:m_b^2}{\al s - m_c^2}, \\ && \nn\\
  \al_{min} &=& \frac{\lambda_{bc}}{2} (1 - v), \\&& \nn \\
  \al_{max} &=& \frac{\lambda_{bc}}{2} (1 + v) ~~.
\enqa

\end{document}